\documentclass[preprint]{vgtc}               

\graphicspath{{figures/}{pictures/}{images/}{./}} 

\usepackage{times}                     

\usepackage{tabu}                      
\usepackage{booktabs}                  
\usepackage{lipsum}                    
\usepackage{mwe}                       
\usepackage{csquotes}
\usepackage{mathptmx}                  
\usepackage{amsmath} 

\newcommand{\VersionA}{3DGS version}
\newcommand{\VersionB}{Mesh version}
\onlineid{0}

\vgtccategory{Research}

\vgtcinsertpkg

\title{3D Gaussian Splatting and Mesh-Based Digital Twins: An Exploratory Study for Virtual Reality Tourism}

\author{Maximilian Warsinke\thanks{e-mail: warsinke@tu-berlin.de}\\ %
    \parbox{1.6in}{\scriptsize \centering Quality and Usability Lab, Technische Universität Berlin, Berlin, Germany}
\and Francesco Vona\thanks{e-mail: Francesco.Vona@hshl.de}\\ %
    \parbox{1.6in}{\scriptsize \centering Immersive Reality Lab, Hochschule Hamm-Lippstadt, Lippstadt, Germany}
\and Abm Tariqul Islam \thanks{e-mail: tariqul.islam@digitaltwin.technology}\\ %
    \parbox{1.6in}{\scriptsize \centering DigitalTwin Technology GmbH, Köln, Germany}
\and Tanja Kojić\thanks{e-mail: tanja.kojic@tu-berlin.de}\\ %
     \parbox{1.6in}{\scriptsize \centering Quality and Usability Lab, Technische Universität Berlin, Berlin, Germany}
\and Jan-Niklas Voigt-Antons\thanks{e-mail: jan-niklas.voigt-antons@hshl.de}\\ %
    \parbox{1.6in}{\scriptsize \centering Immersive Reality Lab, Hochschule Hamm-Lippstadt, Lippstadt, Germany}
\and Sebastian Möller\thanks{e-mail: sebastian.moeller@tu-berlin.de}\\ %
    \parbox{1.6in}{\scriptsize \centering Quality and Usability Lab, Technische Universität Berlin, \\ DFKI, Berlin, Germany}}

\teaser{
  \centering
  \includegraphics[width=0.495\linewidth]{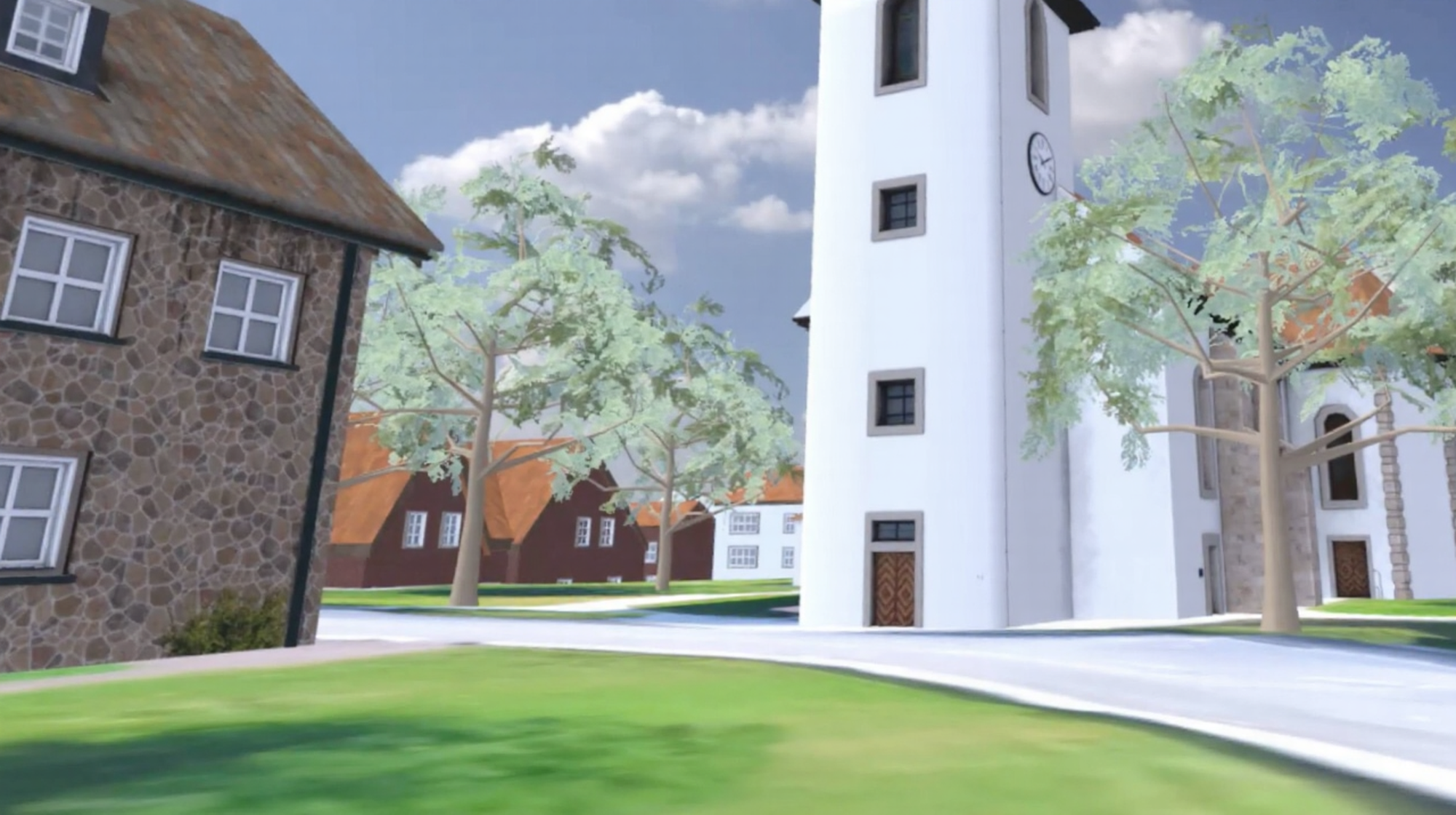}
  \includegraphics[width=0.495\linewidth]{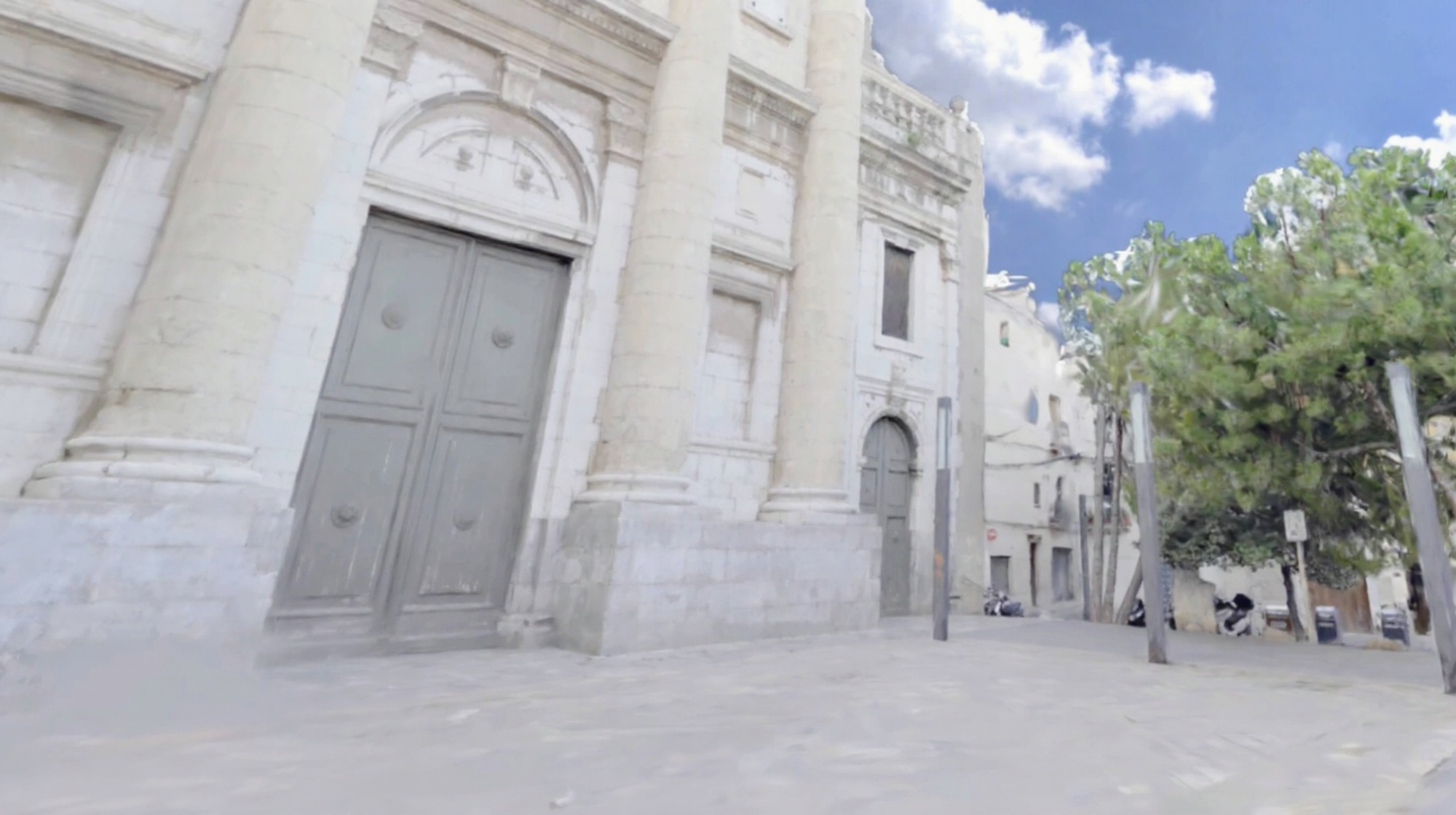}
  \caption{Screenshots of the \VersionB{} (left) and the \VersionA{} (right) of the application.}
  \label{fig:teaser}
}

\abstract{
Digital Twins (DTs) are increasingly used for immersive experiences in virtual tourism. Virtual Reality (VR) enables remote visits to replicated locations for promotional purposes or access to fragile and rural cultural heritage sites. However, developing high-fidelity DTs of tourist destinations is costly, due to the manual creation of 3D environments. Novel 3D rendering techniques, such as 3D Gaussian splatting (3DGS), pose a promising approach to creating immersive experiences. This study investigates the user experience (UX) of a 3D-mesh-based scene and a 3DGS-based scene within a VR tourism application. In a laboratory study, 20 participants engaged with both versions and rated UX, cybersickness, presence and affect through standardized questionnaires. A custom questionnaire was created to measure the perception of the DTs. The collected data suggests that both versions were enjoyed and induced positive affect, with the Mesh version receiving good UX ratings. While the 3DGS version scored higher in terms of experienced realism, it showed clear weaknesses in pragmatic quality. Further, the results suggest that the feeling of presence could be enhanced and cybersickness reduced in both versions. Overall, the study contributes to the understanding of UX in VR tourism applications by implementing mesh-based and 3DGS-based DTs and raising important questions about the perception of realism.
}


\begin{document}



\maketitle


\section{Introduction}

The increased interest in using immersive technology in the tourism and cultural heritage sectors can be summarized as virtual tourism \cite{rifqi2025,tromp2025}. It encompasses enriching on-site experiences with augmented reality (AR) technology and remote exploration through virtual reality (VR) \cite{loureiro2020,baker2023}. The main aspect of VR is its immersive capability, which gives users the sensation of \enquote{being there}, leading to a positive evaluation of the destination \cite{tussyadiah2018}. VR tourism can be used either for promotional purposes to foster travel intentions or to replace a physical visit entirely \cite{wang2026}. For example, it can be used to visit fragile or rural historical sites or as an inclusive, barrier-free opportunity \cite{sang2022,stankov2024}.

The digitally replicated cities, buildings, and sites are more frequently referred to as digital twins (DTs). While the vision behind DTs is the synchronization of physical and digital entities through a continuous, bidirectional data stream \cite{kritzinger2018}, current implementations in VR tourism are often static or unidirectional \cite{almeida2025}. A convergence of VR and DTs is observable due to the synergy between the immersive VR interaction and the practical functionalities of DT technology \cite{tang2025,girginova2026,pandey2025}.

DTs in immersive experiences require high-fidelity visualizations to provide users with an authentic experience \cite{tromp2025}. Therefore, a substantial bottleneck is the visual reconstruction of tourist places and cultural heritage sites, which is currently predominantly achieved through a combination of laser scanning and photogrammetry \cite{pavelkajr2025}. For real-time VR applications, these captures require manual post-processing to clean up the 3D models, which increases costs and development time.

However, novel reconstruction methods, such as 3D Gaussian splatting (3DGS), could be promising for fast and scalable immersive experiences \cite{kerbl2023,wang2025,wang2026}. These approaches allow for high-fidelity 3D content generation from a set of 2D images. On a technical level, these visualization approaches differ fundamentally from polygonal representations, such as traditional 3D-meshes \cite{li2024}. This raises the question of how 3DGS are perceived by users and how they shape the user experience (UX) \cite{wang2026}. Important areas for investigation include cybersickness, realism perception, and sense of presence \cite{li2025}. 

To investigate these dynamics, we developed and evaluated a VR tourism application in a two-part usability study. The application was designed as a demonstrator for remote pre-visit exploration and included two DTs of real locations: Etteln in Germany and Vilanova i la Geltrú in Spain. Both locations were captured using drone imagery but were presented through different reconstruction approaches. Etteln was implemented as a polygonal mesh-based environment, while Vilanova i la Geltrú was reconstructed using 3DGS. Participants explored both versions in VR and rated their experiences using standardized questionnaires covering affect, cybersickness, presence, and UX, complemented by a custom questionnaire assessing perceived DT qualities. By gathering user data from the two versions with a focus on ecological validity, the study contributes to the understanding of UX dimensions in VR tourism and gathers preliminary data on the perception of 3DGS.

\section{Related Work}

\subsection{Virtual Reality Tourism}
While interest in virtual tourism is rising, the idea of using VR and AR technology in tourism contexts has been around for decades. Loureiro et al. provide a comprehensive overview of the usage of VR and AR technology in tourism research that dates back to 1995 \cite{loureiro2020}. They discuss the application of VR in planning, managing, promoting, educating, and creating or transforming tourist experiences. In a more recent review, Rifqi observes an increase in VR tourism studies in 2018, which steadily grew and peaked in 2024 \cite{rifqi2025}. He suggests that this may have been driven by the pandemic, the emergence of AI, and the metaverse.
Tromp et al. review 10 use case studies in the intersection of XR and cultural heritage for similar purposes (e.g., pre-visit planning, on-site exploration) \cite{tromp2025}. They identified six key topics as positive effects while also discussing their challenges: enhanced engagement, broader accessibility, economic benefits for local communities, preservation and promotion, educational impact, and sustainability.
In the broader context of the metaverse, Baker et al. investigated UX factors for virtual heritage tourism \cite{baker2023}. They identified five positive factors, such as presence and authenticity, and eight negative factors, among others: convenience, comfort, and resolution. Additionally, they identified three major dilemmas: the higher costs of more sophisticated hardware and content, the risk of cybersickness from more immersive experiences, and the greater engagement and complexity of gamification.
There is empirical evidence on the benefits of using VR to foster travel intentions. In a two-part study with over 900 participants, Tussyadiah et al. found that feeling present in VR and enjoyment both lead to a positive attitude change, which, in turn, increases the intention to travel \cite{tussyadiah2018}.
Anwar et al. conducted a user study comparing a VR condition to 2D on a smartphone \cite{anwar2025}. Participants experienced a gamified replication of the UNESCO World Heritage site Jabal Al-Ahmar. The results indicate that the VR version outperformed the smartphone version in UX and presence metrics, underlining the benefits of immersive experiences. The presented studies show the viability digital experience in the tourism domain.

%

\subsection{Digital Twins}
The concept of a DT originated in product lifecycle management and has been applied to almost all disciplines and industries \cite{wuni2025}. While there are many definitions, a DT is generally a digital replica of a physical object, place, or system that is synchronized via bidirectional data flow. Kritzinger et al. proposed a classification of DTs with different levels of integration: a digital model (DM) for the absence of data flow, a digital shadow (DS) for unidirectional data flow, and a digital twin (DT) for bidirectional data flow \cite{kritzinger2018}. In a recent review of DT definitions, Wuni et al. rejected this classification and proposed that bidirectional data flow should be a strict requirement, fostering a consistent and cohesive definition \cite{wuni2025}. Despite this critique, this study adopts the DT term as an umbrella concept to facilitate discussion of different levels of digital-physical integration within a unified framework.
According to Tang et al., the convergence of DTs and the metaverse is evident and is driven by AI and extended reality (XR) \cite{tang2025}. This would allow for persistent virtual worlds, enhanced by Internet of Things (IoT) data and perceivable through VR headsets. Girginova describes a co-evolution in which DTs could provide the usefulness that popular XR technologies often lack \cite{girginova2026}, and Pandey et al. focus on use cases in the convergence of DT and XR \cite{pandey2025}. The authors identified manufacturing, healthcare, education, urban planning, and entertainment as the areas that would benefit most from the symbiosis of DTs and XR. Almeida et al. focus on the literature on DTs in the tourism industry and also discuss XR use cases \cite{almeida2025}. They observe the trend that the current DT integration level in virtual tourism is mainly static or unidirectional.
A prime example was presented in a study by Pekridou et al., who built an XR DT application with the highest level of integration (bidirectional data flow) \cite{pekridou2025}. Participants could interact with energy, temperature, and $CO_2$ metrics in real time, enabling human-in-the-loop control. Subjective user data indicated an excellent UX, underlining high engagement, acceptance, and usability. In summary, XR has the potential to serve as an interface to interact with DTs.

\subsection{3D Representations}

There are different ways to visualize 3D content in a virtual scene, with polygonal 3D models being the most widespread. In the broader context of 3D representations, 3D-meshes are categorized as explicit representations, along with voxels and point clouds \cite{liu2024}. Another type of representation is the implicit representation, such as neural radiance fields (NeRF). Explicit representations have clearly defined geometry, while implicit representations utilize mathematical functions to describe their properties \cite{li2024}. 3DGS combines those properties, working similarly to NeRFs but with explicit 3D scene representations \cite{kerbl2023}.
In cultural heritage, authentic 3D reconstructions are achieved through a combination of photogrammetry and laser scanning \cite{bekele2018}. While these approaches produce high-fidelity meshes, they are too complex for real-time applications and require extensive post-processing, such as retopology and texture mapping. This approach was also used by Pavelkajr et al., who first captured the Old Town Bridge Tower in Prague and then optimized the mesh for use in VR and AR \cite{pavelkajr2025}. Using normal maps, depth perception can be mimicked while reducing the number of polygons.
Using 3DGS for 3D representation is expected to become increasingly widespread \cite{wang2025a}. Li et al. reviewed the literature on the usage of radiance fields (NeRF and 3DGS) within the field of XR \cite{li2025}. They discuss opportunities and challenges across topics such as content generation and optimization, and advocate conducting in-depth research on user-centric benchmarks for quality perception, presence, and cybersickness.
Wang et al. compared 3DGS to traditional 3D modeling and 360-degree panoramic imagery in a user study in the context of virtual tourism \cite{wang2026}. Their findings reveal that 3DGS outperformed the other conditions across all dimensions, including realism, emotional arousal, immersion, information credibility, cognitive appraisal, and travel intention. Vachha et al. developed Dreamcrafter, a tool for the ad hoc generation and editing of 3D radiance field representations in VR \cite{vachha2024}. The application was evaluated in a user study in which participants generated and edited mostly 3DGS using voice input and a user interface. Tu et al. identified three key challenges of 3DGS in VR: artifacts, distortions, and reduced frame rates \cite{tu2025}. They tackled those topics with their VRSplat solution, which they validated in a user study.
To conclude, novel 3D representation methods are being evaluated with users and have already been applied in the field of virtual tourism.

\subsection{Research Gap}
The reviewed literature points to a growing convergence of DTs and XR technologies using novel 3D rendering techniques. This is highly relevant for VR tourism: DTs can extend immersive experiences through contextual and potentially real-time information, while emerging 3D reconstruction methods, such as 3DGS, offer a fast, scalable way to generate visually rich 3D environments. While there is ongoing research in the virtual tourism area with mesh-based approaches, 3DGS representations are still largely underexplored. As a result, it remains unclear what implications 3DGS-based DTs can have on UX and perception in real-world use cases.

\section{Methodology}
\subsection{Study Design}
The study tested the two versions of the application in a within-subjects design with repeated measurements. The order of the versions was varied using an AB/BA design to minimize order effects. The study was approved by the ethics committee of [removed for double-blind review]. The study was conducted in a controlled laboratory environment at [removed for double-blind]. The experiment took around 30 minutes, and participants received [removed for double-blind] in compensation.

\subsection{Application}
The application ran on an external PC tethered to a Meta Quest 3 headset via a Meta Link Cable. The PC specifications were an NVIDIA GeForce GTX 1080, an Intel(R) Core(TM) i7-7700K CPU @ 4.20GHz, and 16 GB RAM. Participants interacted with the virtual environments using the standard Meta Quest 3 controllers.
The application was developed as part of the [removed for double-blind review] project using Unity 6. User locomotion was implemented using the Unity XR Interaction Toolkit. Two locomotion modes were enabled: continuous movement and teleportation. Continuous movement allows the user to navigate the environment using the joystick on the left controller, while teleportation is controlled via the right controller joystick and repositions the user to a target location by pointing in the environment. To reduce potential motion sickness, a screen fade (\enquote{blink}) effect is applied during teleportation transitions. Over an interface, users could teleport directly to different points of interest (POIs). In the main menu, two unique versions of the application could be selected.

\subsubsection{3D Gaussian Splatting Version}
The 3DGS version consists of a DT of Vilanova i la Geltrú, generated using novel 3D reconstruction techniques and rendered using a 3DGS representation (see \cref{fig:teaser}). Visual data were collected using drone-based image acquisition. Camera poses and sparse scene geometry were first estimated using Structure-from-Motion (SfM), after which the scene was optimized into a representation composed of 3D Gaussian primitives. This representation approximates the radiance and geometry of the environment, enabling efficient real-time rendering of the reconstructed scene. In this version, the environment consists of several isolated POIs, such as the Church of Sant Antoni and Plaça dels Lledoners. Each scene is reconstructed and rendered independently, rather than integrated into a single spatially continuous urban model. Consequently, users are not able to freely traverse the city environment. Instead, navigation between POIs is performed via an interactive menu provided within the application interface. The DT of Vilanova i la Geltrú can be classified as a digital model in terms of integration.

\subsubsection{3D Mesh Version}
The \VersionB{} consists of a DT of Etteln, generated through a drone imagery-based mesh generation pipeline combined with neural radiance field-based reconstruction techniques (see \cref{fig:teaser}). High-overlap multi-view drone images were first used to optimize a NeRF model for each selected POI. An explicit surface mesh was subsequently extracted from the learned density field using iso-surface extraction (e.g., Marching Cubes) and further optimized to produce a high-fidelity mesh representation suitable for real-time rendering. Key landmarks were further refined through post-processing steps, including tools such as 3ds Max. These enhancements included geometry corrections, higher levels of detail, and improved texture quality to increase visual realism and semantic recognizability within the DT.
In contrast to the \VersionA{}, the mesh-based reconstruction produces a continuous, spatially coherent virtual environment. This enables users to freely explore the village in a natural way while still allowing teleportation between POIs if desired. All POIs are embedded within the unified spatial model. Additionally, selected POIs include information panels that display real-time data from IoT sensors in Etteln, such as the windmill's charging power. The level of integration of the Etteln DT is therefore in the digital shadow category (unidirectional data flow).


\subsection{Metrics}

\begin{table}
  \centering
    \caption{Metrics and questions of the DT questionnaire.}
  \begin{tabular}{l p{5cm}}
  \hline
    Metric & Question \\
    \hline
    Enjoyment     & ``How much did you enjoy the experience?'' \\
    Understanding & ``The application has helped me better understand a place or a system.'' \\
    Information   & ``The application provides information that would otherwise be difficult or impossible to access.'' \\
    Reliability   & ``The application reliably reflects the current status of a place or system.'' \\
    Simulation    & ``The application allows me to simulate processes or scenarios.'' \\
    Comments & ``Additional comments (optional).'' \\
    \hline
  \end{tabular}
  \label{tab:DT_questions}
\end{table}

The demographic questionnaire was designed to collect basic information about participants, including gender, age, and profession. Additionally, their expertise was assessed using three questions targeting different relevant categories: experience with XR, experience with 3D modeling, and video game experience. The reason for including the expertise was the possibility of conducting correlation analysis with the dependent variables.

The Affinity for Technological Interaction (ATI) scale was used to assess participants' general tech affinity \cite{franke2019}. The questionnaire includes questions regarding the willingness and motivation to engage with novel technical systems. Nine Likert scale answers must be provided in the range of 1 to 6. The average ATI score helps to estimate how tech-prone the sample group is, while individual scores can be used in a correlation analysis.

The Affective Slider \cite{betella2016} was employed to measure arousal and valence. The authors suggest using at least 100 steps to maintain slider continuity. Due to the questionnaire format, a 1 to 10 scale was used.

The short version of the User Experience Questionnaire was used to assess the pragmatic quality and hedonic quality of the experience \cite{schrepp2017}. Eight word pairs must be rated on a 1 to 7 scale. Pragmatic and hedonic quality scores both consist of four questions, averaged and converted to a $-$3 to 3 scale. From both sub-dimensions, an overall quality score is derived. The short version was chosen to reduce the length of the questionnaire and avoid overwhelming participants.

Cybersickness was measured using the Cybersickness in Virtual Reality Questionnaire (CSQ-VR) \cite{CSQ-VR}, which consists of the dimensions of nausea, oculomotor, and vestibular. Each symptom score is averaged from two questions, each answered on a 1 to 7 scale. Additionally, participants are given the opportunity to describe their symptoms in detail through qualitative answers. The individual scores are not averaged but summed, which results in a scale from 2 to 14 for the subdimensions and from 6 to 42 for the total cybersickness score.

The sense of being present in the virtual environment was measured using the Igroup Presence Questionnaire (IPQ) \cite{ipq}. It is the most widely adopted questionnaire for this purpose in the literature \cite{Tran2024} and encompasses four dimensions: general presence, spatial presence, involvement, and experienced realism. Questions are rated on a scale of 1 to 7 with varying anchors. Responses were later rescaled to match a -3 to 3 scale for benchmark comparisons.

A custom questionnaire was created to evaluate the perceived DT qualities, as no comparable instrument exists to our knowledge. To date, DT-focused applications are evaluated through common questionnaires from the field of Human-Computer Interaction. To gain more insight into the added value from DTs, four main properties were derived from the literature: The ability of the DT to help users understand a place or system, whether there was information that would otherwise be impossible to access, how up-to-date the DT seemed to participants, and finally, the ability of simulation. Additionally, overall enjoyment of the experience was added, along with the option to provide additional open-ended comments. The exact formulation used can be found in \cref{tab:DT_questions}. A 1 to 7 scale was used with Likert scale anchors.

\subsection{Procedure}

Before starting the study, participants signed the informed consent form and received a brief explanation of the goals and the experiment's structure. They were asked to fill in the demographics questionnaire and the ATI. Each participant was then assigned a starting location and began exploring either the \VersionB{} or the \VersionA{} for a total duration of five minutes. During this exploration phase, they were free to navigate the environment and interact with the available elements.
After completing the first session, participants removed the VR headset and were guided to the first round of questionnaires. Once the questionnaires were completed, the participants returned to the VR headset and repeated the same procedure at the second location. A final set of questionnaires was completed after exploring the second environment, concluding the study.

\subsection{Participants}
A total of 20 participants took part in the study (9 male, 11 female). The mean age of the sample was 33.8 years (SD = 11.31). Participants reported moderate prior experience with XR technologies of 4.05 out of 7 (SD = 1.57), a lower experience with 3D video gaming of 2.85 (SD = 2.32), and the lowest experience was reported with 3D modeling of 2.6 (SD = 1.41). The mean ATI score was 4.27 (SD = 0.88). All participants were recruited through the [removed for double-blind review] participant portal.

\subsection{Data Analysis}
The questionnaire responses were scored according to the scoring guidelines, which involved the reversal of several question items and the conversion from a 1 to 7 scale to a -3 to 3 scale for both UEQ-S and IPQ responses. Descriptive statistics were calculated using JASP \cite{JASP2025}. The standardized questionnaire scores were treated as continuous variables, while the responses to the custom questions were treated as ordinal variables, as they rely on single Likert Scale responses. Therefore, both the median and the arithmetic mean are included in \cref{tab:descriptiveStatisticsA} and \cref{tab:descriptiveStatisticsB}.

The user factors assessed through the demographic questionnaire, specifically prior experience, age, and ATI, were tested against the dependent variables using a correlation analysis. A non-parametric approach was chosen because the prior experience measures were ordinal. Therefore, Spearman rank correlation tests were applied using Python and \texttt{scipy.stats} \cite{SciPy}. To account for Type I errors through multiple comparisons, Holm–Bonferroni corrections were applied to control the family-wise error rate using \texttt{statsmodels.stats.multitest} \cite{statsmodel}. Each of the five user factors was treated as a separate family, containing the correlations between that score and the 14 dependent variables. For brevity, only remaining statistically significant correlations are reported in the results section. A significance threshold of $\alpha < 0.05$ was used to determine statistical significance. No statistical tests were conducted to compare the two versions, as the study design did not isolate a single variable.

\section{Results}
Presented are descriptive statistics of the questionnaire responses and the results of correlations between user factors and dependent variables within each version.

\begin{table}[t]
    \centering
    \caption{Descriptive statistics for the \VersionA{}.}
    \label{tab:descriptiveStatisticsA}
    \begin{tabular}{lrrr}
        \toprule
         Metric & Median & Mean & SD \\
        \midrule
          Valence & 7.00 & 5.95 & 2.63 \\
         Arousal & 6.00 & 6.10 & 2.34 \\
        \hline
         Enjoyment & 5.00 & 4.40 & 1.88 \\
        Understanding & 4.50 & 3.95 & 1.70 \\
        Information & 2.50 & 3.15 & 2.01 \\
        Reliability & 4.00 & 3.55 & 1.43 \\
        Simulation & 4.50 & 4.30 & 1.69 \\
        \hline
        G. Presence & 0.50 & 0.50 & 1.61 \\
        Exp. Realism & -0.63 & -0.71 & 1.06 \\
        Involvement & 0.25 & 0.13 & 1.52 \\
        Sp. Presence & 0.30 & 0.07 & 1.45 \\
        \hline
        Pragmatic Q. & 0.25 & 0.41 & 1.36 \\
        Hedonic Q. & 0.88 & 0.61 & 1.58 \\
        \hline
        Cybersickness & 17.00 & 19.50 & 9.76 \\
        \bottomrule
    \end{tabular}
\end{table}

\begin{table}[t]
    \centering
    \caption{Descriptive statistics for the \VersionB{}.}
    \label{tab:descriptiveStatisticsB}
    \begin{tabular}{lrrr}
        \toprule
        Metric & Median & Mean & SD \\
        \midrule
        Valence & 7.00 & 6.80 & 1.77 \\
        Arousal & 7.00 & 7.10 & 1.48 \\
        \hline
        Enjoyment & 5.00 & 5.35 & 1.50 \\
        Understanding & 5.00 & 5.05 & 1.43 \\
        Information & 4.00 & 4.15 & 1.76 \\
        Reliability & 4.50 & 4.30 & 1.38 \\
        Simulation & 5.50 & 5.40 & 1.31 \\
        \hline
        G. Presence & 1.00 & 1.05 & 1.50 \\
        Exp. Realism & -1.00 & -0.83 & 0.96 \\
        Involvement & 0.38 & 0.51 & 1.21 \\
        Sp. Presence & 0.60 & 0.66 & 1.05 \\
        \hline
        Pragmatic Q. & 1.88 & 1.70 & 0.90 \\
        Hedonic Q. & 1.13 & 1.16 & 1.19 \\
        \hline
        Cybersickness & 16.00 & 17.80 & 8.92 \\
        \bottomrule
    \end{tabular}
\end{table}

\subsection{\VersionA{}}
The following presents the mean scores of the dependent variables gathered through post-session questionnaires for the \VersionA{}. The full descriptive statistics can be found in \cref{tab:descriptiveStatisticsA}.

In terms of affect, participants reported a mean Valence score of 5.95 and a slightly higher average Arousal score of 6.10.
The UEQ-S scores show a mean Pragmatic Quality of 0.41, a higher Hedonic Quality score of 0.61, resulting in an average Overall score of 0.51 (SD = 1.31). The total IPQ score was -0.39 (SD = 0.97).
Participants reported a mean Cybersickness score of 19.50 as scored by the CSQ-VR. For the individual sub-dimensions, Vestibular symptoms were reported as the strongest, with a score of 6.75 (SD = 3.77), followed by Nausea with 6.65 (SD = 3.45) and Oculomotor with 6.10 (SD = 3.71). The qualitative responses highlighted blurry and unclear visuals, flickering, and unnatural locomotion, leading to nausea and discomfort during the experience.
The IPQ results showed the highest scores for General Presence at 0.50, followed by Involvement at 0.13 and Spatial Presence at 0.07. Experienced Realism received the lowest score at -0.71. 
The custom DT Questionnaire showed the best mean ratings in the category Enjoyment with 4.40, followed by Simulation with 4.30. Understanding received a score of 3.95, and Reliability scored 3.55. Lastly, Information received the lowest rating with 3.15.

In the \VersionA{}, one significant correlation between user factors and outcome measures remained after applying the Holm–Bonferroni correction. Prior Experience with 3D modeling correlated with perceived Reliability ($r_s = 0.68$, $p < 0.001$, $p_{\mathrm{Holm}} = 0.012$).

\subsection{\VersionB{}}
The following presents the average scores of the dependent variables for the \VersionB{}. Comprehensive descriptive statistics are presented in \cref{tab:descriptiveStatisticsB}.

Participants reported a mean Valence score of 6.80 and a slightly higher average Arousal score of 7.10.
The UEQ-S scores indicate a higher mean Pragmatic Quality of 1.70 compared to a Hedonic Quality score of 1.17. On average, this results in an Overall UEQ-S score of 1.43 (SD = 0.87).
Participants reported a mean Cybersickness score of 17.80. The individual dimensions were rated as follows: Oculomotor was highest at 6.50 (SD = 3.09), followed by Nausea at 6.10 (SD = 3.78) and Vestibular at 5.20 (SD = 3.53). Qualitative responses attributed symptoms of motion sickness and dizziness primarily to fast movement, snap-view transitions, and navigating confined spaces, such as the windmill. Discomfort was also associated with moving in VR while remaining stationary in the real world, although some participants reported little to no symptoms.
In terms of presence scores as assessed by the IPQ, General Presence scores highest with 1.05, followed by Spatial Presence with 0.66 and involvement with 0.51. The lowest score was given to Experienced Realism with an average score of -0.83. This resulted in a total IPQ score of -0.06 (SD = 0.70).
Regarding DT Perception, Simulation received the highest score of 5.40, closely followed by Enjoyment with 5.35. Understanding was rated 5.05 on average, Reliability was rated 4.30, and lastly, Information Access was rated 4.15.

In the \VersionB{}, one significant correlation remained between user factors and outcome measures after applying the Holm-Bonferroni correction. Prior Experience with 3D modeling correlated with Experience Realism ($r_s = 0.65$, $p = 0.002$, $p_{\mathrm{Holm}} = 0.024$).


\section{Discussion}
Both arousal and valence showed scores above average in both versions, indicating a positive, arousing experience. In terms of UX, the \VersionB{} received a higher score in pragmatic quality compared to hedonic quality, indicating that users favored usefulness and usability over enjoyment and appeal. Compared to the official UEQ-S benchmark\footnote{https://www.ueq-online.org/}, pragmatic quality was rated ``Good'', whereas hedonic quality was only ``Above Average''. The total score is still located in the ``Good'' category, indicating an overall successful implementation of the use case from the user's perspective. The scores of the \VersionA{} only reached the ``Below Average'' category in terms of hedonic quality and the ``Bad'' category in pragmatic quality, showing a lack of usability and usefulness in this version. The overall UEQ-S score is also considered ``Bad'' according to the benchmark, highlighting the negative impact of the usability aspects. Apart from the visual differences, the \VersionB{} offered a higher degree of interaction through real-time sensor data integration and an interconnected world, which likely improved the UX.

Cybersickness was reported in both versions at similar low to average level compared to the scales' full range. However, a few users reported severe symptoms, which could pose a challenge for usage in virtual tourism. Interestingly, the qualitative responses revealed insights into potential causes. In the \VersionB{}, participants reported discomfort mainly from locomotion, while in the \VersionA{}, visual artifacts seemed to have caused symptoms. One explanation could be that the \VersionB{} with a large virtual environment encouraged more movement. Furthermore, visual disturbances from 3DGS are known to threaten UX in VR \cite{tu2025}.

\begin{figure}[t]
    \centering
    \includegraphics[width=1\linewidth]{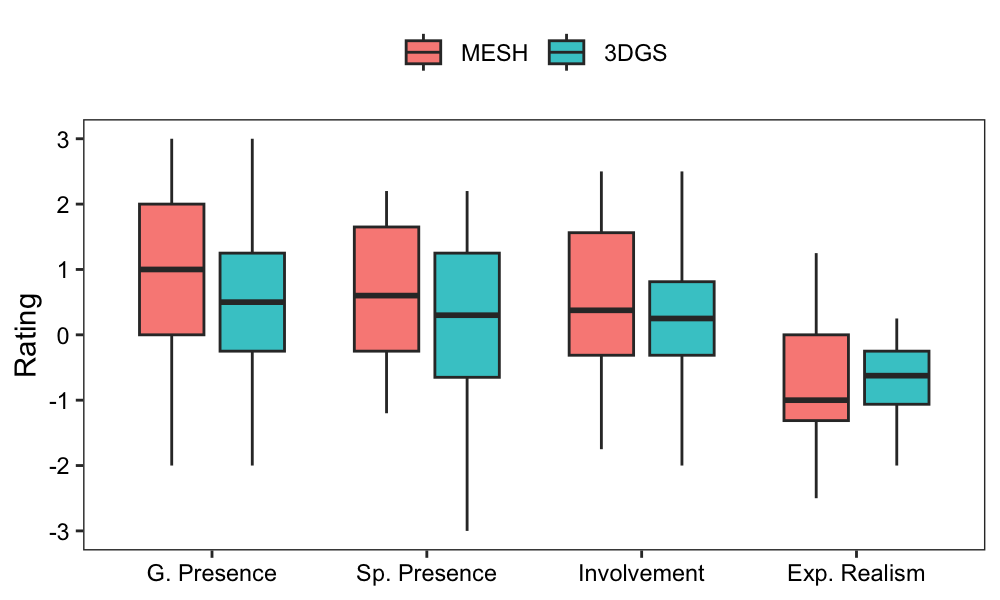}
    \caption{Boxplots showing the distributions of IPQ responses for both versions.}
    \label{fig:boxplot}
\end{figure}

\begin{figure}[t]
    \centering
    \includegraphics[width=0.95\linewidth]{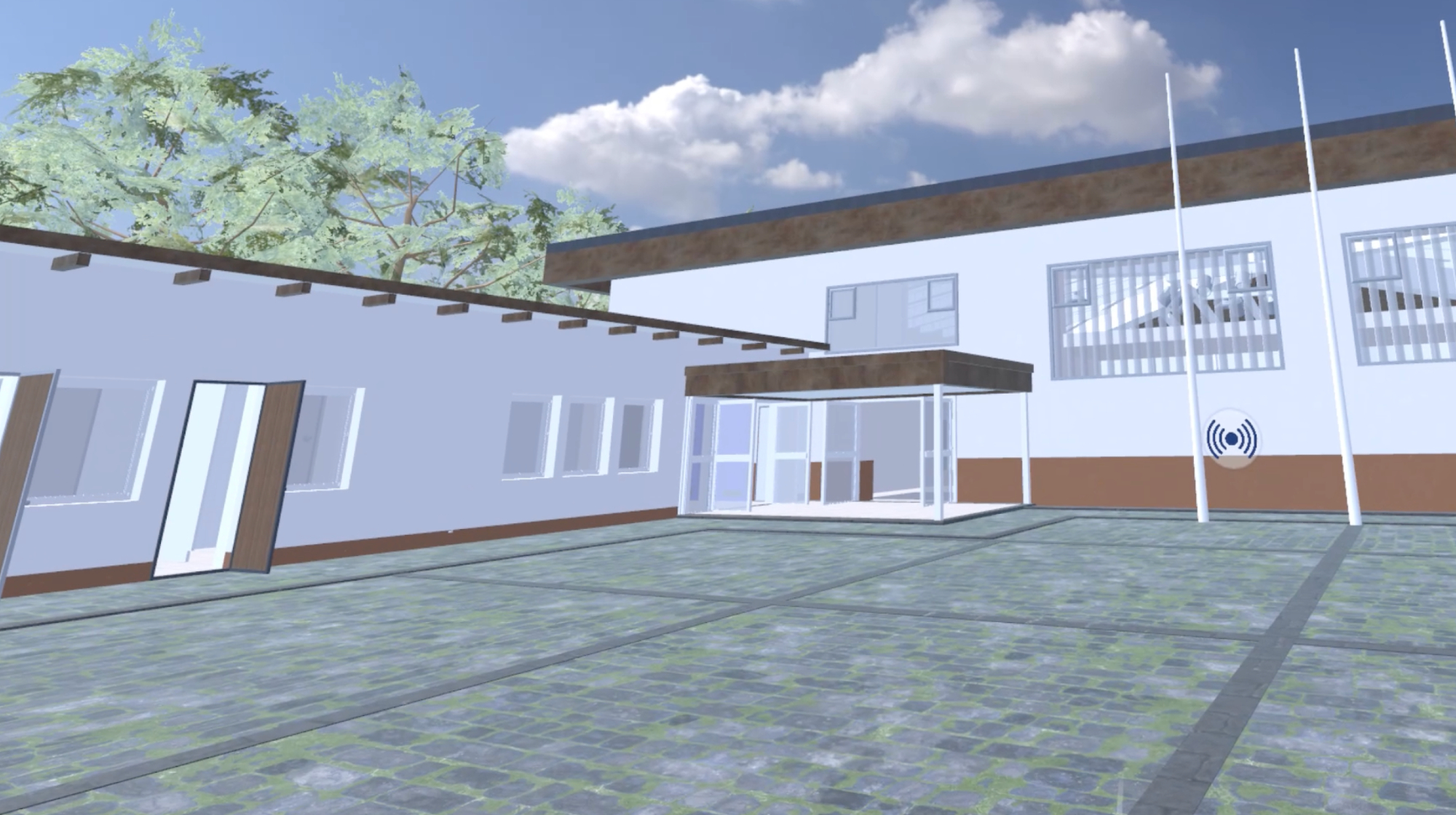}
    \caption{Screenshot from the \VersionB{} that depicts the simplified 3D-mesh geometry of a building.}
    \label{fig:etteln_detail}
\end{figure}

\begin{figure}[t]
    \centering
    \includegraphics[width=0.95\linewidth]{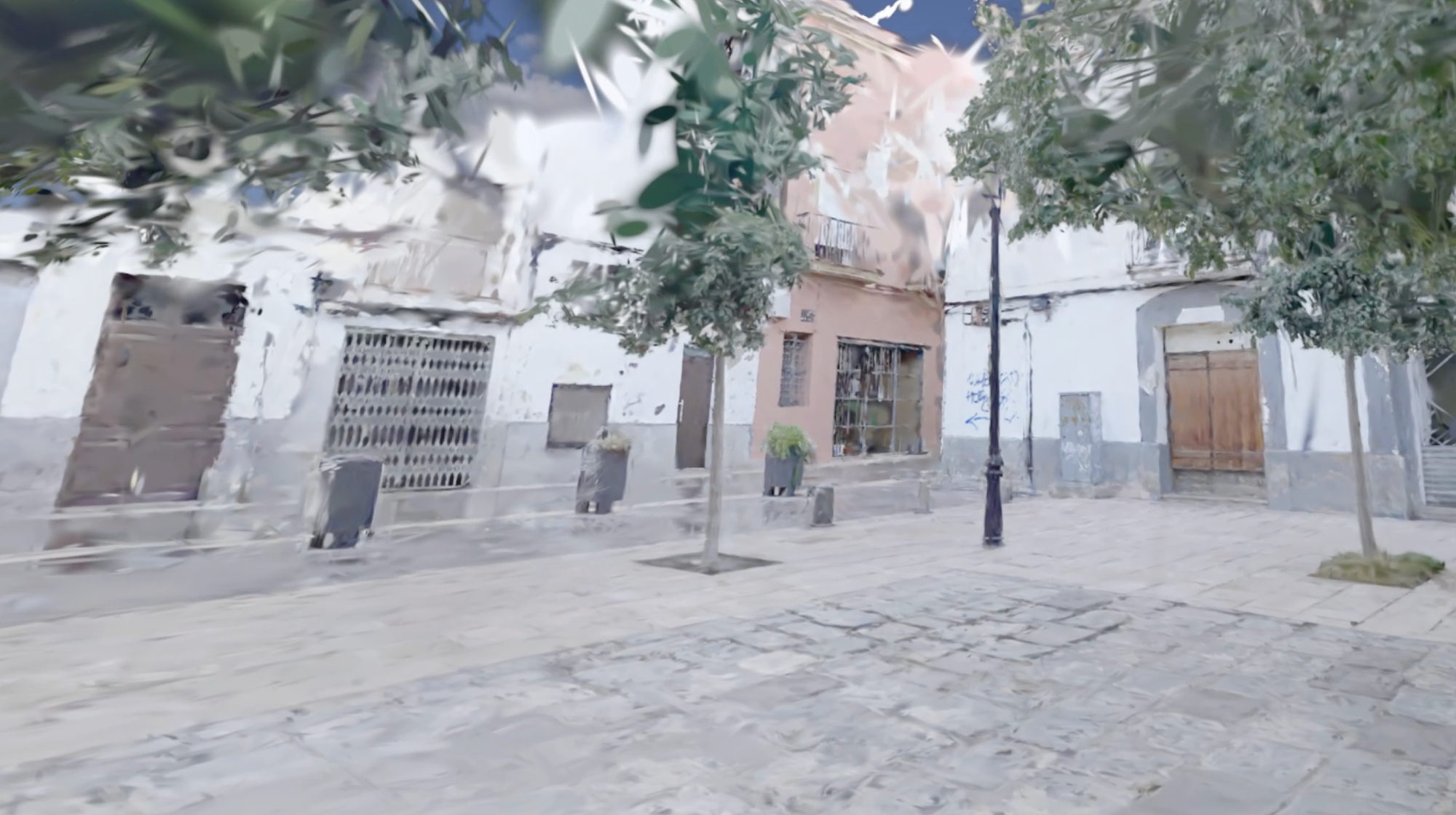}
    \caption{Screenshot from the \VersionA{} that shows rendering artifacts prominent in the treetops.}
    \label{fig:vilanova_detail}
\end{figure}

The presence scores were overall low compared to the benchmark from Tran et al. \cite{Tran2024}. In the \VersionB{}, the Involvement and General Presence scores reached the ``Moderate'' threshold, while the other dimensions received a ``Low'' score. In the \VersionA{} all scores were in the ``Low'' category. Interestingly, Experienced Realism is the only metric, apart from cybersickness, that scored higher in \VersionA{} compared to the \VersionB{}, likely due to the rendering technique that has an overall more photorealistic appeal (see \cref{fig:boxplot}). Still, the IPQ scores were unexpectedly low in both versions. A two-fold explanation can be considered that reflects two opposite perception dynamics from the rendering techniques. The appearance of the environment in the \VersionB{} can be classified as low-fidelity, as the point cloud data was reduced to simple mesh geometry during the creation process. \cref{fig:etteln_detail} exemplifies the appearance of the buildings with a low level of detail, both in texture and polygon count. On the contrary, the \VersionA{} has a more detailed appearance, with a enhanced level of detail and high resolution ``textures''. However, the scene has more inconsistencies and imperfect areas with visual artifacts, such as the treetops shown in \cref{fig:vilanova_detail}. Therefore, both approaches could have resulted in low presence and realism scores for their unique reasons.

An attempt to quantify the perception of the DTs was made through the custom questionnaire created for the study. As the question items are not validated, the interpretations shouldn't be overestimated. Enjoyment, Simulation, and Understanding scored the highest, while Reliability and Information received lower scores in both versions. This aligns with the capabilities of the application, whose aim is to simulate a visit to a destination of choice. Apparently, the experience helped the understanding but did not necessarily provide additional information. In that regard, it must be noted that the sensor data in the \VersionB{} was possibly not seen by each participant, as they could explore freely without mandatory content. While the geometry was not updated further after the initial DTs were created, users still reported a moderate level of reliability regarding the current status of the places' representation.


The sample group was evenly distributed by gender and spanned a diverse age range. The average affinity for technology interaction was above average (4.27 out of 6), indicating a technologically interested group. Their prior XR experience was average, while their video game and 3D modeling experiences were low. Compared to the average tourist, the sample group may be slightly more technologically advanced but was still considered suitable for the study.

In the \VersionB{}, prior 3D modeling experience showed a moderate positive association with experienced realism. This may indicate that users with greater expertise in 3D modeling evaluated the visuals as more realistic, possibly due to a greater awareness of the technical constraints and nuances of 3D content creation. 3D modeling is commonly understood as the manipulation of mesh geometry, which could explain why the relationship was not observed in the \VersionA{}. In the \VersionA{} 3D modeling experience and reliability showed a moderate positive correlation. Reliability in that regard refers to the ability to reflect the current status of a place or system (see \cref{tab:DT_questions}). A possible explanation is that participants with more 3D modeling experience were more familiar with how physical environments are translated into digital models. Consequently, they may have been more confident in the correspondence between the reference location and its DT. However, this interpretation remains speculative, as it is unclear why the same relationship was not observed in the \VersionB{}.


\subsection{Limitations and Future Work}
Some limitations of the study should be acknowledged. Most importantly, the study was not designed as a controlled experiment with a single manipulated variable. Instead, multiple factors varied simultaneously between the two versions, such as the replicated place or the connectedness of the environment. Consequently, we refrained from performing inferential statistical comparisons between the versions and avoided drawing direct conclusions about their relative performance. Any comparative statements are therefore descriptive only. Additionally, participants could freely explore the application within a given time. Users visited the POIs in a different sequence and for varying durations. Generally, the study's sample size was small, and the results of the DT questionnaire should be interpreted with caution, as it is not standardized. The questionnaire was used solely because, to our knowledge, no validated instrument currently exists for the subjective assessment of DT perception.

Future work should investigate the exact implications of 3D representation approaches on UX. To gain more nuanced insights, subsequent studies should focus on a subset of metrics and employ a more controlled study design. One approach would be to create DTs of the same environments using different 3D representation methods in a 2-by-2 crossover design to systematically investigate the effect of 3DGS compared to more traditional reconstruction approaches. Furthermore, the discussed differences in visual fidelity and their effects on experienced realism could be promising for efficient resource allocation in DT creation process. Another path could focus on the DT integration aspect by updating the environment or adding bidirectional data flow. Lastly, metrics measuring usefulness to the tourism industry, such as travel intention, should be considered for evaluating the application on a use case level.

\section{Conclusion}
The study examined two versions of an application for VR tourism in terms of UX, presence, affect, cybersickness, and attempted to quantify DT perception through a custom questionnaire. The \VersionB{} with an interconnected virtual environment and IoT sensor integration was well received, with good UX ratings and low to moderate presence scores. The \VersionA{} with POI-centric exploration and 3DGS rendering showed weaknesses in pragmatic quality and low presence scores. Interestingly, experienced realism received a higher score in the \VersionA{}, which may be attributed to the rendering technique. This indicates that increased realism perception alone is not necessarily associated with an increased UX. While cybersickness occurred in both versions, participants reported movement-related issues in the \VersionB{}, whereas in the \VersionA{}, rendering artifacts caused discomfort. Despite symptoms, the VR experience was perceived positively, as indicated by affect and enjoyment ratings. The custom questionnaire responses indicate the strength of simulating the tourist destinations, which improved the understanding of the places. The exploration of user factors revealed that higher 3D modeling experience was positively associated with both perceived realism (\VersionB{}) and perceived reliability (\VersionA{}). Overall, the user evaluation suggests that both versions could benefit from higher presence and that the functionalities of the \VersionB{} likely enhanced pragmatic quality. The results contribute to the understanding of UX in VR tourism and provide preliminary insights into the perception of emerging 3D reconstruction methods. 

\section{Acknowlegements}
This work was supported by the European Union’s Horizon Research and Innovation Program under Grant 101092875 (DIDYMOS-XR: Digital DynaMic and responsible twinS for XR). The Overleaf AI features and Grammarly were used for editing and grammar enhancement during the document writing process.

\bibliographystyle{abbrv}

\bibliography{main}
\end{document}